# 4G Wireless Networks: Opportunities and Challenges

Hassan Gobjuka
Verizon
919 hidden Ridge
Irving, TX 75038

*Abstract*
**With the major wireless service providers planning to start deployment of 4G wireless networks by mid 2010, research and industry communities are racing against time to find solutions for some of the prominent still open issues in 4G networks. The growing interest in 4G networks is driven by the set of new services will be made available for the first time such as accessing the Internet anytime from anywhere, global roaming, and wider support for multimedia applications.**
**In this paper describe some of the key opportunities will be made available by 4G networks, present key challenges and point to some proposed solutions.**

## I. INTRODUCTION

The existence of 4G Networks in today's technology-driven society is important indicators of advancement and change. 4G, or Fourth Generation networks, are designed to facilitate improved wireless capabilities, network speeds, and visual technologies. It is anticipated that as these networks continue to thrive, the demand for advanced related technologies will also grow, thereby creating new alternatives for savvy technology users to exceed their desired expectations. The following discussion will evaluate the current state of 3G Networks and will examine the future potential of these networks in expanding technology-based capabilities for consumers and industries alike.

In this paper we present an overall vision of the 4G networks starting by presenting some of the key features they will provide, and then discussing key challenges the researchers and vendors are attempting to resolve, and finally briefly describing some of the proposed solutions to these problems.

## II. BACKGROUND

Within the cable television industry, the expansion to 4G Networks is a very real possibility in 2009. Recently, Comcast and T-Mobile have collaborated and agreed to the development of a "mobile 4G network" to be tested in Washington D.C. and Baltimore, MD [11]. However, this process is lengthy, and the rollout of such a network is not expected for close to two years, as the network requires extensive and detailed testing in order to ensure that there are no "bugs" that could interrupt the flow of mobile traffic across the network [11]. This type of opportunity is of critical importance in developing a network that is capable of advancing technology to never-before-seen heights.

Similarly, AT&T, one of the world's largest telecommunications providers, will begin its own rollout of a 4G Network in 2011, enabling its vast user base to explore new downloading speeds and capabilities [18]. The utilization of LTE mobile broadband technology is an opportunity for the corporation to expand its horizons into 4G territory, upstaging current 3G capabilities [18]. In the process of expanding into the new 4G enterprise, AT&T will seek to overcome any limitations brought on by the 3G Network process. As AT&T begins its rollout process, there are many considerations involved in ensuring that the transition is a success, and that existing networks are not interrupted in the process of developing the 4G platform. In addition, mobile providers such as AT&T will likely develop new pricing strategies from some of their most popular products, including the iPhone, in response to the challenges of developing faster networks [21].

The 4G Network process requires a unique approach to developing effective models for strategic purposes. The necessity for 4G networks is associated with the increased utilization of data websites such as You Tube and Facebook, which require tremendous bandwidth in order to be used successfully [10]. Because these websites are becoming increasingly popular amongst the general public, it is more important than ever for telecommunications providers to develop opportunities to accommodate the needs of the consumer population. Consumers have come to depend on different sources of data as a source of entertainment and for convenience. Therefore, it is important that organizations such as Verizon and AT&T continue to identify areas where technological improvements are required.

In January 2009, the first operating 4G Network was established by a joint venture between Clearwire and Intel, which reflected an opportunity for residents and businesses in Portland, Oregon to "connect wirelessly anywhere in Portland at true broadband speeds" [13]. However, with the technology quickly approaching a widespread rollout, many cities, states, and countries will soon possess similar capabilities, as consumers and businesses alike will be provided with different opportunities to expand their networks and interfaces with advanced capabilities. Furthermore, it is evident that the Clearwire strategy is not without its disadvantages, and additional efforts must be made to overcome any technology-related problems that might persist before a widespread rollout is even considered.

### III. OPPERTUNITIES

In general, it is believed that the existence of the 4G network is designed to facilitate the development of a superior alternative to the existing 3G strategy in terms of quality and data transmission speed. For developers of 4G Networks, there is a great dependence upon advanced technologies and increased speed in order for the network to be a success. It is known that in terms of the 4G Network, "it requires substantial improvements to multimedia messaging services, including video services, in order to approve a new generation. It wants a data speed transfer rate of at least 100 megabits per second while a user is physically moving at high speeds and a one gigabit per second data rate in a fixed position" [5]. From this perspective, it is important for the new data network to meet the expected demand of the consumer and of different industries, which have come to depend upon high-speed data networks with minimal interruptions for a variety of needs.

*A. Cost and Affordability*

In terms of 4G Network cost and affordability, there are a number of issues to consider that reflect some degree of risk, as well as opportunity, so that these networks are successful once rolled out to the general public, and in general, 4G Networks are designed in order to create an environment that supports high-speed data transmission and increased profit margins for organizations that utilize these capabilities [2]. Developing a successful 4G Network platform is a positive step towards the creation of a wireless and broadband environment that possesses rapid transmission speeds, data integrity modules, and other related events that encourage users to take additional risks in promoting successful utilization of these 4G tools.

*B. Cababilities and Features*

Although the 4G Network platform is not brand new, many telecommunications providers have not yet developed their own alternatives that will support this network in full. Therefore, 4G-related products are still in the development phase, with additional products to be developed and rolled out on a periodic basis. With the creation of these alternatives, it is likely that 4G Networks will continue to expand their scope and promote their own brand of personalization for consumers that seek these types of alternatives [16]. In general, the possibilities associated with 4G Networks are endless, as high-speed data transmission and associated capabilities are more feasible than ever. This supports the notion that the demand for more complex networks and related capabilities are stronger than ever, as a greater number of consumers continue to buy into the potential that exists with advanced networks, such as 4G.

With the appropriate combination of resources, it is possible for 4G Networks to create alternatives that exceed consumer and industry expectations. Therefore, 4G developers must consider the appropriate security measures, the promotion of high-speed data transmission across the network, and must also consider the ways in which data quality and integrity might be preserved in order to provide the most satisfactory results.

This 4G is intended to replace the current 3G systems within few years. The ambitious goal of 4G is to allow everyone to access the Internet anytime and everywhere. The provided connection to Internet will allow users to access all type of services including text, databases, and multimedia. 4G, unlike 3G, is IP based, that is every user connected to the Internet will have an IP address. This feature makes it easier to integrate the infrastructure of all current networks and consequently will it easier for users to access services and applications regardless of the environment. 4G will also provide higher bandwidth, data rate, lower authentication overhead, and will ensure the service is constantly provided to the user without any disruption.

Another key feature of 4G networks is high level of user-level customization. That is, each user can choose the preferred level of quality of service, radio environment, etc. Accessing 4G networks will be possible virtually by using any wireless device such as PDAs, cell phones, and laptops. Figure 1 illustrates elements and techniques to support the adaptability of the 4G domain.

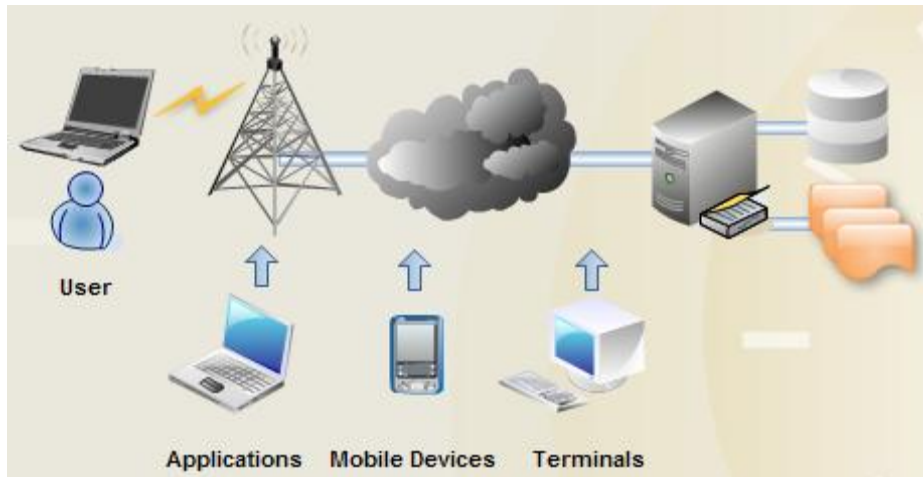
**Figure 1:** 4G will allow everyone to access the Internet from everywhere using almost any wireless device.

## IV. NEW CHALLENGES

### A. Security and Privacy

In the development of 4G Networks, security measures must be established that enable data transmission to be as safe as possible. Specifically, "The 4G core addresses mobility, security, and QoS through reuse of existing mechanisms while still trying to work on some mobility and handover issues" [3]. Therefore, it is necessary for the organization to develop an effective series of tools that support maximum 4G security measures as a means of protecting data that is transmitted across the network from hackers and other security violations. Because of the nature of the 4G network, there is an increased likelihood of security attacks, and therefore, multiple levels of security, including increased requirements for authentication, will be necessary to protect data and information that is transmitted across the network [3].

One of the main goals of G4 networks is to blanket very wide geographic area with seamless service. Obviously, smaller local area networks will run different operating systems. The heterogeneity of these wireless networks exchanging different types of data complicates the security and privacy issues. Furthermore, the encryption and decryption methods being used for 3G networks are not appropriate for 4G networks as new devices and services are introduced for the first time in 4G networks. To overcome these security and privacy issues, two approaches can be followed. The first is to modify the existing security and privacy methods so that they will be applicable to heterogeneous 4G networks. Another approach is to develop new dynamic reconfigurable, adaptive, and lightweight mechanisms whenever the currently utilized methods cannot be adapted to 4G networks [14].

### B. Quality of Service

With respect to network quality, many telecommunications providers are promising that there will be enhanced connectivity, and the quality of data that is transmitted across the network will be of the highest possible quality, as in the case of Ericsson's 4G Network for TeliaSonera [7]. The company promises that "The new 4G network will do for broadband what mobile telephony did for voice. With real-time performance, and about 10 times higher data rates compared to today's mobile broadband networks, consumers can always be connected, even on the move" [7]. As a result, it is important for providers to develop an effective approach to the 4G Network that will enhance quality, provide effective security measures, and will ensure that all users are provided with extensive alternatives for downloading video, music, and picture files without delays.

The main challenge that 4G networks are facing is integrating non-IP-based and IP-based devices. It is known that devices that are not IP address based are generally used for services such as VoIP. On the other hand, devices that are IP address based are used for data delivery. 4G networks will serve both types of devices. Consequently, integrating the mechanisms of providing services to both non-IP-based as well as IP-based devices is one of key challenges 4G networks have to address [17, 19].

## C. Complex Architecture

### C.1. Multimode End-User Terminals

To reduce operating costs, devices that operate on 4G networks should have the capability to operate in different networks. This will not only reduce the operating cost but will also simplify design problems and will reduce power consumption. However, accessing different mobile and wireless networks simultaneously is one of the major issues 4G networks have been addressing. One mechanism that has been proposed to handle this problem is termed "multi-mode devices". This mechanism can be achieved through a software radio that allows the end-user device to adapt itself to various wireless interfaces of the networks. Figure 2 shows an example of such solution.

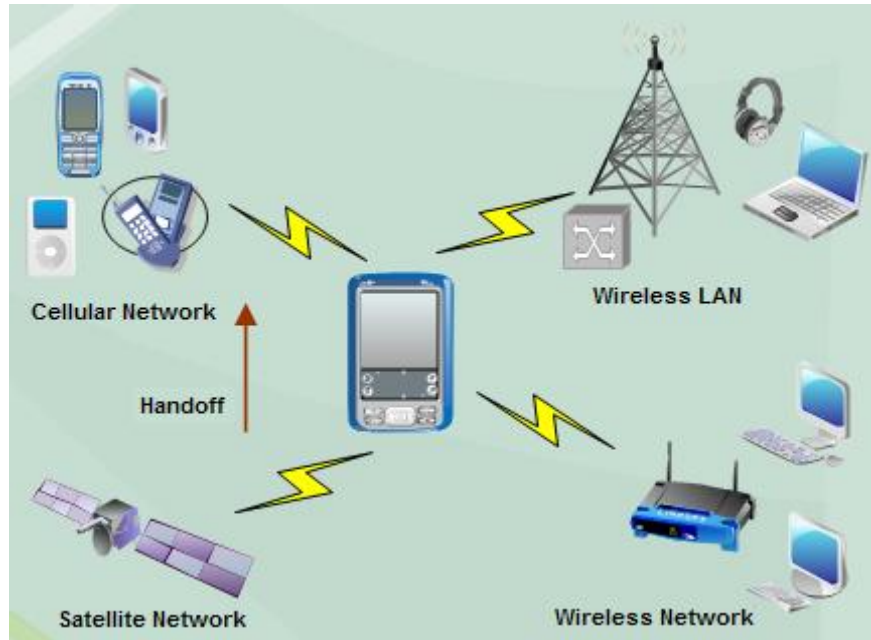

**Figure 2:** Accessing multiple networks and services through multi-mode software

### C.2. System Discovery and Selection

Due to the heterogeneity of 4G networks, wireless devices have to process signals sent from different systems, discover available services, and connect to appropriate service providers. Various service providers have their own protocols which can be incompatible with each other as well as with the user's device. This issue may complicate the process of selecting the most appropriate technology based on the time, place and service provided, and thus, may affect the Quality of service provided to the end user.

One solution to resolve this issue is called "System-initiated discoveries". This mechanism allows automatic download of software modules based on the wireless system the user is connected to [12]. Another approach to handle this problem is based overlay networks. In such case, the end-user device is connected to different networks through an overlay network. The overlay network performs all necessary tasks such as protocol translation and Quality of service negotiation as depicted in Figure 3.

### C.3. Service and Billing

Managing user accounts and billing them has become much more complicated with 4G networks. This is mainly due to heterogeneity of 4G networks and the frequent interaction of service providers. The research community addressed this concern and proposed several frameworks to handle the customers' billing and user account information [8, 9].

## V. CONCLUSION

4G wireless networks not only enable more efficient, scalable, and reliable wireless services but also provides wider variety of services. These opportunities come with a need for rethinking our security, privacy, architect and billing technologies have been used for previous generations. We believe, however, that future research will overcome these challenges and integrate newly developed services to 4G networks making them available to everyone, anytime and everywhere.

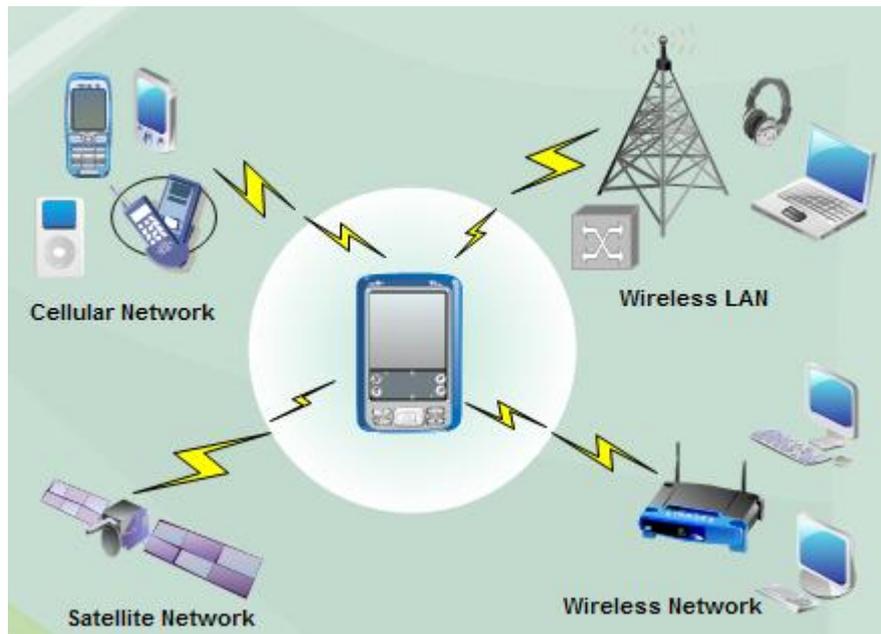

**Figure 3:** Automatic system discovery is one of the features provided by 4G networks.

## REFERENCES

[1] 3GPP TS 23.107 v.5.9.0, "Quality of Service (QoS) Concept and Architecture," June 2003.

[2] Accenture, Accenture 4G accelerator solution. 2009.

[3] E. Buracchini, "The Software Radio Concept," IEEE Commun. Mag., vol. 38, no. 9, 2000, pp. 138–43.

[4] J. Chavis, What is a 4G network?, 2009

[5] H. Eguchi, M. Nakajima, and G. Wu, "Signaling Schemes over a Dedicated Wireless Signaling System in the Heterogeneous Network," Proc. IEEE VTC, Spring 2002, pp. 464–67.

[6] Ericsson, Ericsson to build commercial 4G network for TeliaSonera. 2009.

[7] J. Fleck, "A Distributed Near Real-time Billing Environment," Telecommun. Info. Net. Architecture, 1999, pp. 142–48.

[8] F. Ghys and A. Vaaraniemi, "Component-based Charging in a Next-generation Multimedia Network," IEEE Commun. Mag., vol. 41, no. 1, Jan. 2003, pp. 99–102.

[9] S. Higgenbotham, Countdown to 4G: who's doing what, when, 2008.

[10] R. Jackson, T-Mobile 4G network coming with help from Comcast, 2009.

[11] T. H. Le and A. H. Aghvami, "Performance of an Accessing and Allocation Scheme for the Download Channel in Software Radio," Proc. IEEE Wireless Commun. And Net. Conf., vol. 2, 2000, pp. 517–21.

[12] A. Lyle, Clear, first 4G network launched, 2009

[13] J. Al-Muhtadi, D. Mickunas, and R. Campbell, "A Lightweight Reconfigurable Security Mechanism for 3G/4G Mobile Devices," IEEE Wireless Commun., vol. 9, no. 2, Apr. 2002, pp. 60–65.

[14] N. Montavont and T. Noel, "Handover Management for Mobile Nodes in IPv6 Networks," IEEE Commun. Mag., vol. 40, no. 8, Aug. 2002, pp. 38–43.

[15] D. Perkins, 4G: challenges and opportunities, 2007.

[16] A. D. Stefano and C. Santoro, "NetChaser: Agent Support for Personal Mobility," IEEE Internet Comp., vol. 4, no. 2, Mar./Apr. 2000, pp. 74–79.

[17] P. Taylor, AT&T to roll out 4G network, 2009.

[18] B. Thai and A. Seneviratne, "IPMoA: Integrated Personal Mobility Architecture," Comp. and Commun., 2001, pp. 485–90.

[19] D. Tipper et al., "Providing Fault Tolerance in Wireless Access Networks," IEEE Commun. Mag., vol. 40, no. 1, Jan. 2002, pp. 58–64.

[20] S. Wildstrom, AT&T's Stephenson: the road to 4G, 2009.